\documentclass{article}
\usepackage{fullpage,amsmath,amsfonts,setspace,cancel,url,hyperref,verbatim, cite}
\newcommand{\be}{\begin{equation}}
\newcommand{\ee}{\end{equation}}

\newcommand{\tX}{\tilde X}
\newcommand{\tY}{\tilde Y}
\newcommand{\tZ}{\tilde Z}
\newcommand{\talpha}{\tilde \alpha}
\newcommand{\tbeta}{\tilde \beta}
\newcommand{\sumn}{\sum_{n\neq 0}}

\numberwithin{equation}{section}
\begin{document}

\begin{titlepage}
\title{
Non-commutativity and non-associativity of the doubled string in non-geometric backgrounds}
\author{Chris D. A. Blair 
\\ \\ 
 Department of Applied Mathematics and Theoretical Physics,\\ Centre for Mathematical Sciences, University of Cambridge,\\
Wilberforce Road, Cambridge, CB3 0WA, United Kingdom
\\
\\
Email: \url{c.d.a.blair@damtp.cam.ac.uk}
}
\date{ }
\maketitle

\begin{abstract}
We use a T-duality invariant action to investigate the behaviour of a string in non-geometric backgrounds, where there
is a non-trivial global $O(D,D)$ patching or monodromy. This action leads to a set of Dirac brackets
describing the dynamics of the doubled string, with these brackets determined only by the monodromy. 
This allows for a simple derivation of non-commutativity and non-associativity in backgrounds which
are (even locally) non-geometric. We focus here on the example of the three-torus with H-flux, 
finding non-commutativity but not non-associativity. 
We also comment on the relation to the exotic $5^2_2$ brane, which shares the same monodromy. 
\end{abstract} 
\thispagestyle{empty}
\end{titlepage}

\newpage 

\tableofcontents

\section{Introduction}

Dualities lie at the heart of much of our understanding of string and M-theory. However, the normal
formulations of these theories may not be the best suited for an efficient description of these
dualities, and may not allow us to fully appreciate their effects. As a result, one is led to the
idea of modifying and extending the original formulations, to obtain duality manifest models. 

In this paper we will be concerned with T-duality in string theory, where such attempts were first pioneered some time ago 
\cite{Duff:1989tf, Tseytlin:1990nb, Tseytlin:1990va, Siegel:1993th, Siegel:1993xq}.
The basic idea to be followed is the extension of spacetime by introducing dual coordinates. This leads to
worldsheet sigma model descriptions of the string moving in a doubled space
\cite{ 
Duff:1989tf,Tseytlin:1990nb, Tseytlin:1990va,
 Hull:2004in,Dabholkar:2005ve, 
Hull:2006va, Hull:2006qs, 
Hull:2007jy, Hull:2009sg,
Berman:2007xn,West:2010rv,
Maharana:2010sp,Maharana:2012fv,
Copland:2011yh,Copland:2011wx,
Nibbelink:2012jb, 
Lee:2013hma,
Blair:2013noa,
DeAngelis:2013wba,
 Plauschinn:2013wta
}. 
Similarly, supergravity can be rewritten using the ideas
of generalised geometry
\cite{
Hitchin:2004ut, Gualtieri:2003dx
}
leading to duality manifest formulations
\cite{
Coimbra:2011nw, Coimbra:2012yy,
Siegel:1993th, Siegel:1993xq,
Hull:2009mi,
Hull:2009zb, 
Hohm:2010pp, 
Hohm:2010xe,
Hohm:2011zr, 
Jeon:2011cn,
Jeon:2012hp
} including double field theory, which is defined on a space with twice as many coordinates as
usual. Double field theory can be thought of as the low energy effective theory of the doubled string
\cite{Berman:2007xn,
Copland:2011yh,Copland:2011wx}, while the latter in turn can be seen to reappear by studying
fluctuations of the string solution in double field theory \cite{Berkeley:2014nza}. 
For recent reviews of these theories, see \cite{Aldazabal:2013sca, Berman:2013eva, Hohm:2013bwa}. 

One of the ambitions of these approaches has been to understand so-called ``non-geometric backgrounds.''
 \cite{Hull:2004in, Dabholkar:2005ve,
Grana:2008yw, 
Kachru:2002sk,Hellerman:2002ax,Flournoy:2004vn,Shelton:2005cf}.
These are backgrounds which are patched by T-dualities rather than just normal diffeomorphisms and
gauge transformations; they are important in the context of flux
compactifications\cite{Shelton:2005cf}. By considering such backgrounds on the same stringy footing
as any other we are able to extend the set of branes we have to play with to include new ``exotic''
objects related by T-dualities to familiar branes \cite{deBoer:2010ud, deBoer:2012ma}. The
non-geometrical features of these backgrounds may be better understood using ideas related to the
doubled formalism such as non-geometric frames
\cite{
Andriot:2011uh, Andriot:2012wx,Andriot:2012an,Andriot:2013xca,
Blumenhagen:2012pc,Blumenhagen:2012nk,Blumenhagen:2012nt,Blumenhagen:2013aia,
Blumenhagen:2013hva,
Hassler:2013wsa,
Geissbuhler:2013uka,
Andriot:2014uda 
}, while the ``patching'' by a T-duality can be understood as a particular finite
coordinate transformation of the doubled geometry \cite{Hohm:2012gk,Hohm:2013bwa, Park:2013mpa, Berman:2014jba,
Papadopoulos:2014mxa}.

The behaviour of strings in such backgrounds is expected to deviate from our usual expectations. In particular, there is plenty of evidence pointing towards non-commutativity and non-associativity in the non-geometric setting
\cite{ 
Lust:2010iy,Blumenhagen:2010hj,Blumenhagen:2011ph,Condeescu:2012sp, Mylonas:2012pg, Andriot:2012vb, Bakas:2013jwa
}. In this paper our aim is to use the doubled string action developed in \cite{Blair:2013noa} to
examine this behaviour. The advantage of this approach is that T-duality is manifest, and it is
trivial to treat in a unified manner non-geometric backgrounds and their (geometric)
T-duals\footnote{There is also a notion of a ``truly non-geometric'' background which would not be
connected by duality to a geometric background, for example see \cite{Hassler:2014sba}. It should be
possible to treat these in similar fashion using our methods, however we will not consider any such
case in the present paper.}. Indeed from the point of view of the doubled space the notion of a
non-geometric background does not make sense: all such backgrounds are (generalised) geometric. Only
when we single out half of the coordinates as belonging to a physical spacetime do we have a
meaningful notion of non-geometric properties. 

We will focus mainly on the three-torus with H-flux. Although not a true string theory background,
this provides an interesting simplified model of the sort of non-geometric effects we are interested
in, and the behaviour under T-duality of the string in this background has received plenty of
attention
\cite{
Lust:2010iy,
Blumenhagen:2011ph,
Mylonas:2012pg, Andriot:2012vb, Bakas:2013jwa,
DallAgata:2008qz, 
Halmagyi:2008dr, Halmagyi:2009te, 
Hull:2009sg, 
Rennecke:2014sca
}.
By carrying out T-dualities
along the two isometry directions of this model one obtains firstly a twisted torus (with geometric
flux) and secondly a non-geometric background (with Q-flux). In the non-geometric background the metric and B-field can
only be globally defined up to a T-duality transformation. A further T-duality along a direction
which is not an isometry leads to a background which depends explicitly on a dual coordinate, and so
is not even locally geometric (though it may still be thought of as carrying an R-flux).

In the
non-geometric background one encounters non-commutativity of string coordinates, and we will be able
to give a simple derivation of the explicit non-commutative bracket found in \cite{Andriot:2012vb}. There a detailed
and careful study of the T-duality rules was used to carry results from the geometric backgrounds over to the non-geometric one: we will be able to do a much more straightforward computation of 
the Dirac brackets of the doubled model leading to the same result 
(note that our results are all classical and in terms of Poisson or Dirac brackets, but this carries directly over to the quantum theory).

Using the doubled model also gives us more control over the situation in which the background is not
even locally geometric. By treating coordinates and their duals on an equal footing there is no
difficulty at least in principle with considering T-dualities along directions which are not
isometries. Thus we are also able to work out explicitly the brackets in this case. These
brackets are more pathological.
However by computing the Jacobi identity, which vanishes, we find that in fact our brackets do not display the
non-associative property. This may be due to a subtle discrepancy between the coordinates used here
and those used in other investigations such as \cite{Blumenhagen:2011ph}.

In our doubled model the only piece of information about the background which is relevant to
determining the Dirac brackets and hence non-commutativity (or non-associativity) is the $O(D,D)$ monodromy which describes the (lack of) periodicity of the generalised metric (which in the T-duality description unifies the metric and B-field). This is a consequence of the fact that the underlying origin of such effects, 
is the mixing of physical and dual coordinates in the closed string boundary conditions, which is
also encoded by the monodromy. 

This means that although the three-torus with H-flux is not a true string background but rather an interesting toy
model, 
a T-duality chain of genuine string theory backgrounds with the same monodromy would inherit
all our results. A particular example is provided by the NS5-KKM-$5_2^2$ T-duality chain, and one
may conclude (as mentioned in \cite{Hassler:2013wsa}) that a string in the background of the exotic $5_2^2$
brane also exhibits non-commutativity.

The outline of this paper is as follows. In section 2 we review the T-duality chain of the three-torus with H-flux, and write down the
standard string Polyakov action for this background. In section 3 we give the modification of our
doubled action which allows us to describe strings in backgrounds with non-trivial $O(D,D)$
monodromy. We then proceed to use this to study a doubled string in the background defined by the
three-torus with H-flux: this doubled string simultaneously describes the various interesting
T-duals of the original background. We show the equivalence of our description with the usual one.
In section 4 we analyse the Poisson brackets and constraints of the doubled model and work out the
Dirac brackets which are used to describe the dynamics: these Dirac brackets give rise to
non-commutativity in the non-geometric background. In section 5 we then extend our analysis to study
the effects of carrying out a T-duality along a direction which is not an isometry. We calculate here a
triple bracket of string coordinates to 
investigate 
the appearance of non-associativity. Finally we
include a short appendix to demonstrate the similarities between the toy background considered here
and the $5_2^2$ brane. 

\section{Torus with H-flux: ordinary sigma model and dual coordinates}

\subsection{Reminder of T-duality in the doubled formalism}

We will be interested solely in backgrounds with non-vanishing metric $g$ and B-field $B$. The T-duality
properties of these fields can be expressed by combining them into the generalised metric:
\be
\mathcal{H}_{MN} = 
\begin{pmatrix} 
g-Bg^{-1} B & Bg^{-1} \\
- g^{-1} B & g^{-1} 
\end{pmatrix}
 = \begin{pmatrix} 1 & B \\ 0 & 1 \end{pmatrix} \begin{pmatrix} g & 0 \\ 0 & g^{-1} \end{pmatrix} 
\begin{pmatrix} 1 & 0 \\ - B & 1 \end{pmatrix} \,.
\label{eq:genMetricDefn}
\ee
T-duality transformations $P_M{}^N$ are elements of the group $O(D,D)$, and so by definition obey
\be
P_M{}^P \eta_{PQ} (P^T)^Q{}_N = \eta_{MN} \quad , \quad \eta_{MN} = \begin{pmatrix} 0 & 1 \\ 1 & 0 \end{pmatrix} \,.
\ee
The $O(D,D)$ structure $\eta_{MN}$ can be used to raise and lower the indices on the transformation
matrix $P_M{}^N$.

In the doubled picture the coordinates consist of the usual spacetime coordinates $X$ together with their duals $\tilde X$, which form a single $O(D,D)$ vector 
\be
X^M = \begin{pmatrix} X \\ \tilde X \end{pmatrix} \,.
\ee
In order to make contact with the usual spacetime picture we have to specify a polarisation, or choice of coordinates which we take to be the physical ones. One way of seeing the action of T-duality is to keep the polarisation fixed and rotate the geometry:
\be
\bar{X}^M = P^M{}_N X^N \quad , \quad \bar{\cal{H}}_{MN} ( \bar{X} ) = P_M{}^P \mathcal{H}_{PQ} (X)(P^T)^Q{}_N\,.
\ee
The new ``physical'' coordinates after this T-duality will still be the upper $D$ components of $\bar{X}^M$. An equivalent point of view is to keep the geometry (i.e. the generalised metric) fixed but have T-duality act on the polarisation, i.e. after acting with a T-duality we will now select a different set of physical coordinates. 

\subsection{T-dual backgrounds}

Our example background for investigating non-geometric effects is that of the three-torus with
H-flux. It is important to remember that this is not a true string theory background. We will mainly
be interested in viewing it as a toy model with which to develop our understanding, although one can
be more precise about viewing it as an approximation to a true background as for instance in
\cite{Andriot:2012vb}. 

\subsubsection{Torus with H-flux}

We take the metric on the torus to be flat $ds^2 = dX^2 +dY^2 + dZ^2$ and pick a gauge such
that the B-field is $B = HZ dX \wedge dY$. There is then a constant flux $H = H dX \wedge dY \wedge dZ$ through the
three-torus. This background can be treated as a two-torus with coordinates $X,Y$ fibred over a base
circle parameterised by the coordinate $Z$. For the time being we will only double this two-torus,
introducing dual coordinates $\tilde X$ and $\tilde Y$. The generalised metric is then a
four-by-four matrix, and has the form
\be
\mathcal{H}_{MN} = \begin{pmatrix} (1+ H^2 Z^2) \mathbb{I}_2 & \varepsilon HZ \\ - \varepsilon HZ &
\mathbb{I}_2 \end{pmatrix} \quad ,\quad\varepsilon = \begin{pmatrix} 0 & 1 \\ - 1 & 0 \end{pmatrix} \,.
\label{eq:GenMetHFlux} 
\ee
As we loop around the $Z$ direction this generalised metric is not periodic, but changes as
\be
\mathcal{H}_{MN} (Z+2\pi) = P_M{}^P \mathcal{H}_{PQ}(Z) (P^T)^Q{}_P \,,
\ee
with $O(2,2)$ monodromy
\be
P_M{}^{N} = \begin{pmatrix}
\mathbb{I}_2 & 2\pi H \varepsilon \\ 0 & \mathbb{I}_2
\end{pmatrix}  \,.
\ee
This monodromy has a straightforward physical interpretation as a gauge transformation of the
B-field: $B \rightarrow B + d ( 2 \pi H X \wedge dY)$, as is easily seen from the decomposition of the generalised metric in \eqref{eq:genMetricDefn}.

\subsubsection{One T-duality: the twisted torus}

We implement T-duality in the $X$ direction by the matrix
\be
( T_X )^M{}_N = 
\begin{pmatrix}
0 & 0 & 1 & 0 \\
0 & 1 & 0 & 0 \\
1 & 0 & 0 & 0 \\
0 & 0 & 0 & 1 
\end{pmatrix} \,.
\ee
The physical configuration this leads to is known as the twisted torus. The physical coordinates are
$\tilde X$ (the T-dual of the original coordinate $X$), $Y$ and $Z$. This background has vanishing B-field, and metric
\be
ds^2 = ( d \tX - H Z dY )^2 + dY^2 +dZ^2 \quad,\quad \tX \sim \tX + 2 \pi H Y \,.
\ee
We need to make a geometric identification of the coordinates $\tX, Y$ as we go around the $Z$ circle for this space to make sense. However, this is nothing out of the ordinary. It corresponds to an $O(2,2)$ monodromy
\be
P^\prime_M{}^N = 
\begin{pmatrix}
A & 0 \\ 0 & A^{-T} 
\end{pmatrix} \quad ,
\quad A = \begin{pmatrix} 1 & 0 \\ - 2\pi H & 1 \end{pmatrix}\,,
\ee
which lies in the $\mathrm{GL}(2)$ subgroup of $O(2,2)$ corresponding to coordinate transformations.
This background is considered to have geometric flux, $f^i{}_{jk}$ with $f^x{}_{yz} = H$.

\subsubsection{Another T-duality: the non-geometric background}

T-duality in the $Y$ direction corresponds to the following element of $O(2,2)$
\be
( T_Y )^M{}_N = 
\begin{pmatrix}
1 & 0 & 0 & 0 \\
0 & 0 & 0 & 1 \\
0 & 0 & 1 & 0 \\
0 & 1 & 0 & 0 
\end{pmatrix} \,,
\ee
and produces a non-geometric configuration:
\be
ds^2 = \frac{1}{1 + H^2 Z^2} ( d\tX^2 + d\tY^2) +dZ^2 \quad,\quad B = - \frac{HZ}{1+H^2 Z^2} d\tX \wedge d\tY \,.
\ee
Though the generalised metric is still periodic up to an $O(2,2)$ element
\be
P^{\prime \prime}_{M}{}^{N} = \begin{pmatrix}
\mathbb{I}_2 & 0 \\ 2 \pi H \varepsilon & \mathbb{I}_2 
\end{pmatrix} \,,
\ee
this transformation however cannot be interpreted as a coordinate transformation or gauge
transformation. The background is said to carry a Q-flux, $Q^{ij}{}_k$, with $Q^{xy}{}_z = H$. 

\subsubsection{A T-duality too far?}

The conventional B\"uscher T-duality rules \cite{Buscher:1987sk, Buscher:1987qj}, only apply along directions which are isometries. However, in the doubled framework every coordinate appears together with its dual. This allows one to at least attempt to make sense of carrying out T-duality in arbitrary directions. If we were to therefore perform a further T-duality in the $Z$ direction we would end up in a situation where our physical space has coordinates $\tX,\tY$ and $\tZ$, and fields 
\be
ds^2 = \frac{1}{1 + H^2 Z^2} ( d\tX^2 + d\tY^2 ) + d\tZ^2 \quad , \quad B = - \frac{HZ}{1+H^2 Z^2} d\tX \wedge d\tY \,.
\ee
The fields still depend however on $Z$ which is the dual coordinate in this picture. As a result,
this space is believed to not even be locally geometric. However it can still be thought of as
carrying an R-flux, $R^{ijk}$, with $R^{xyz}= H$. 

For the time being we shall ignore the possibility of carrying out this T-duality, and return to it in section \ref{sec:noniso}.

\subsection{The sigma model and dual coordinates}

Our goal in this paper is to use a doubled formalism to study all of the above backgrounds simultaneously. As a check on the doubled formalism we will first set up the usual string sigma model in the simplest case of the flat three-torus with H-flux. 

The Polyakov action in the H-flux background is 
\be
S = \int d\tau d\sigma \left[ \frac{1}{2} \left( \dot{X}^2 + \dot{Y}^2 + \dot{Z}^2 - X^{\prime 2} - Y^{\prime 2} -
Z^{\prime 2} \right) + HZ ( \dot{X} Y^\prime - \dot{Y} X^\prime ) \right] \,,
\label{eq:PolyHFlux}
\ee
leading to the equations of motion
\be
\begin{split}
\ddot{X} - X^{\prime \prime}  - H \dot{Y} Z^\prime + H Y^\prime \dot{Z} & = 0 \,, \\
\ddot{Y} - Y^{\prime \prime} - H \dot{Z} X^\prime + H Z^\prime \dot{X} & = 0 \,,\\
\ddot{Z} - Z^{\prime \prime} - H \dot{X} Y^\prime +  H X^\prime \dot{Y} & = 0 \,.
\label{eq:singlestringeom}
\end{split} 
\ee
The momenta are
\be
P_X \equiv \dot{X} + H Z Y^\prime \quad, \quad
P_Y \equiv \dot{Y} - H Z X^\prime \quad,\quad
P_Z \equiv \dot{Z} \,.
\ee
The action \eqref{eq:PolyHFlux} is equivalent to the Hamiltonian form
\be
S = \int d\tau d\sigma \left[ \dot{X} P_X + \dot{Y} P_Y + \dot{Z} P_Z - \frac{1}{2} \mathcal{H}_{MN} \mathcal{Z}^M \mathcal{Z}^N - \frac{1}{2} (P_Z)^2 - \frac{1}{2} (Z^\prime)^2\right]\,,
\ee
where $\mathcal{H}_{MN}$ is the generalised metric given in \eqref{eq:GenMetHFlux} and
\be
\mathcal{Z}^M = \begin{pmatrix} X^\prime \\ Y^\prime \\ P_X \\ P_Y  \end{pmatrix} \,.
\ee
We want to describe string configurations with winding in all directions, so that $X(\sigma+2\pi) =
X(\sigma) + 2 \pi N^1$ and similarly for the others. This means that we can write mode expansions 
\be
\begin{split}
X & = x(\tau) + N^1 \sigma + \sumn \alpha_n e^{in\sigma}\,, \\
Y & = y(\tau) + N^2 \sigma + \sumn \beta_n e^{in\sigma}\,, \\
Z & = z(\tau) + N^3 \sigma + \sumn \gamma_n e^{in\sigma} \,.
\label{eq:modeexpansions}
\end{split}\ee
While here the winding numbers $N^i$ as constant, in the doubled picture they will be dynamical variables.

The dual coordinates $\tilde X$ and $\tilde Y$ are defined in terms of the momenta by
\be
\tilde X^\prime = P_X = \dot{X} + HZ Y^\prime  \quad,\quad
\tilde Y^\prime = P_Y = \dot{Y} - HZ X^\prime  
\,.
\ee
If we have winding then these are not periodic but obey
\be
P_X(\sigma+2\pi) = P_X(\sigma) + 2\pi HN^3 Y^\prime(\sigma) \quad,\quad
P_Y(\sigma+2\pi) = P_Y(\sigma) - 2\pi HN^3 X^\prime(\sigma) \,.
\ee
Using the mode expansions \eqref{eq:modeexpansions} one can explicitly show that
\be
\begin{split}
\tX & = \tilde x + \left( \dot{x} + H \sumn in \gamma_{-n} \beta_n - HN^3 y + H N^2 z \right) \sigma
+ H N^3 \sigma \left( y + \sumn e^{in\sigma} \beta_n \right) 
+\\ & 
\sumn e^{in \sigma} \frac{1}{in} \left( \dot{\alpha}_n + in H z \beta_n - H N^3 \beta_n 
+ H N^2 \gamma_n 
+  H \sum_{m\neq 0} im \gamma_{n-m} \beta_m \right)
 + \frac{1}{2} H N^2 N^3 \sigma^2 \,.
\end{split}
\label{eq:tXConstr}
\ee
This quantity is also not periodic but obeys
\be
\tX(\sigma+ 2\pi) = \tX (\sigma) + 2 \pi H N^3 Y(\sigma) + 2 \pi p_1\,,
\ee
where 
\be
p_1 = 
\dot{x} -  H N^3 y +  H z N^2 +  \pi H N^3 N^2 +  H \sumn in \gamma_{-n} \beta_n  \,.
\ee
The equations of motion imply that $\dot{p}_1 = 0$.

Similarly, one has
\be
\begin{split}
\tilde Y& = \tilde y + p_2 \sigma - H N^3 \sigma ( X(\sigma,\tau) - \sigma N^1 ) -  \frac{1}{2} HN^3
N^1
\sigma ( \sigma- 
2 \pi )  \\
& + 
\sumn e^{in \sigma} \frac{1}{in} \left( \dot{\beta}_n - in H z \alpha_n + H N^3 \alpha_n - H N^1
\gamma_n -  H
\sum_{m\neq 0} im \gamma_{n-m} \alpha_m \right) \,,
\end{split}
\label{eq:tYConstr}
\ee
with
\be
p_2 = 
\dot{y} +  H N^3 x -  H z N^1 -  \pi H N^3 N^1 -  H \sumn in \gamma_{-n} \alpha_n  \,,
\ee
and
\be
\tY(\sigma+ 2\pi) = \tY (\sigma) - 2 \pi H N^3 X(\sigma) + 2 \pi p_2\,.
\ee
Finally, the momentum conjugate to $Z$ is just $\dot{Z}$, so that
\be
\tilde Z = \tilde z + \dot{z} \sigma + \sum_{n\neq 0} \frac{1}{in} \dot{\gamma}_n e^{in\sigma} \,.
\ee
This has more conventional periodicity properties: 
\be
\tilde Z (\sigma+2\pi) = \tilde Z (\sigma) + 2 \pi p_3 \,,
\ee
where $p_3 = \dot{z}$. As this direction is not an isometry we do not have $\dot{p}_3 = 0$. 

\section{Doubled sigma model for three-torus with H-flux}

\subsection{The action}
\label{sec:actionderivation}
The doubled sigma model will allow us to treat all three of the above backgrounds in a unified
manner. To do so, we need to slightly generalise the derivation of the doubled action from \cite{Blair:2013noa} to describe
non-trivial monodromies. As a first step to doing so, let us recall the steps taken in
\cite{Blair:2013noa} where no non-trivial monodromy was assumed.  

The usual string action in Hamiltonian form and conformal gauge is 
\be
S = \int d \tau d\sigma \left( \dot{X}^i P_i - \frac{1}{2} \mathcal{H}_{MN} \mathcal{Z}^M \mathcal{Z}^N \right) \,.
\label{eq:actionHamForm}
\ee
The Hamiltonian depends only on $X^{\prime i}$ and the momentum $P_i$ in a manifestly $O(D,D)$ invariant form, with the $O(D,D)$ vector
\be
\mathcal{Z}^M = \begin{pmatrix} X^{\prime i} \\ P_i \end{pmatrix} \,.
\ee 
We make the replacement $P_i \mapsto \tilde X^\prime_i$, thereby replacing the momentum with dual coordinates. The kinetic term can be manipulated by integration by parts into the form
\be
\begin{split}
\int d\tau d\sigma \dot{X}^i P_i & = 
\int d\tau d\sigma \frac{1}{2} \dot{X}^M \eta_{MN} X^{\prime N} + \int d\tau d\sigma\frac{1}{2} \frac{d}{d\sigma} \left( \dot{X}^i \tilde X_i \right) \\
& = \int d\tau d\sigma \frac{1}{2} \dot{X}^M \eta_{MN} X^{\prime N} + \pi \int d\tau  \dot{X}^i(0) p_i  \,,
\end{split}
\label{eq:kinetictermmanip}
\ee
assuming for now that we have no $O(D,D)$ monodromy, but do have winding such that $X^i(\sigma+2\pi) = X^i(\sigma)+2\pi w^i$ and $\tilde X_i (\sigma+2\pi) = \tilde X_i + 2\pi p_i$. In the usual string case we have constant $w^i$, but we do not assume this for the momentum zero mode $p_i$. 

Now, the second term of the last line of \eqref{eq:kinetictermmanip} is not $O(D,D)$ covariant. One might proceed naively to drop it from the action entirely in order to obtain a proposed $O(D,D)$ manifest sigma model. However, one can check \cite{Blair:2013noa} that this does not lead to the correct equations of motion or Dirac brackets, with the treatment of the zero modes turning out to be incorrect. The correct solution is to first note that as we are using a Hamiltonian form of the action it involves the dynamical quantity $p_i$, and we wish to treat the winding $w_i$ on the same dynamical footing. One is therefore led to the following modification: adding to the action a term
\be
+ \pi \int d\tau \dot{\tilde{X}}_i (0) w^i \,,
\ee
such that we have total action
\be
S  = \int d\tau d\sigma \left( \frac{1}{2}\dot{X}^M \eta_{MN} X^{\prime N} - \frac{1}{2} X^{\prime M}
\mathcal{H}_{MN} X^{\prime N} \right) + \pi \int d \tau w^M \eta_{MN} \dot{X}^N(0)  \,.
\label{eq:doubledactiontrivmon}
\ee
The generalised winding $w^M$ is treated as dynamical, and must be taken into account when varying
the action. In this way one obtains equations of motion equivalent to those of the Polyakov action,
and the expected covariant Dirac bracket:
\be
\{ X^M(\sigma) , X^N(\sigma^\prime) \}^* = - \eta^{MN} \epsilon(\sigma-\sigma^\prime) \,,
\ee
where $\epsilon^\prime (\sigma) = \delta(\sigma)$, such that we reobtain the usual brackets between
coordinates and momenta. 

The situation is broadly similar for non-trivial monodromies. Given the general boundary condition
\be
X^M(\sigma+2\pi) = P^M{}_N X^N(\sigma) + 2 \pi w^M \,,
\ee
then the generalisation of the action in \cite{Blair:2013noa} which describes non-trivial
monodromies is
\be
S_{doubled} = \int d\tau d\sigma \left( \frac{1}{2}\dot{X}^M \eta_{MN} X^{\prime N} - \frac{1}{2} X^{\prime M}
\mathcal{H}_{MN} X^{\prime N} \right) + \pi \int d \tau w^M \eta_{MN} P^N{}_P  \dot{X}^P(0)  \,.
\label{eq:doubledaction}
\ee
The additional term has the effect of treating the zero modes correctly: it eliminates a cross-term involving $\dot{w}^M$ and the oscillator modes which leads to incorrect Dirac brackets, and ensures the momentum $p_i$ are given by the expression obtained from the usual analysis. 

The equations of motion from the action \eqref{eq:doubledaction} can be expressed as
\be
\begin{split}
& \frac{d}{d \sigma} \left( - \eta_{MN} \dot{X}^N + \mathcal{H}_{MN} X^{\prime N} \right) - \frac{1}{2} \partial_{M} \mathcal{H}_{PQ} X^{\prime P} X^{\prime Q} 
\\ 
& + \delta(\sigma-2\pi) \left( \pi \eta_{MN} \dot{w}^N + P_{MN} \dot{X}^N(0) - \mathcal{H}_{PN} X^{\prime P}(0) P_M{}^N \right) 
\\ &
- \delta(\sigma) \left( \pi P_{NM} \dot{w}^N + \eta_{MN} \dot{X}^N(0) - \mathcal{H}_{NM} (0) X^{\prime N}(0) \right) 
\\
& = 0 \,.
\end{split} 
\ee
Integrating over $\sigma$ from $0$ to $2\pi$ implies that
\be
( \eta_{NM} + P_{NM} ) \dot{w}^N =  - \frac{1}{2\pi} \int_0^{2\pi} d\sigma \partial_{M} \mathcal{H}_{PQ} X^{\prime P} X^{\prime Q} \,.
\ee
If the generalised metric is constant, i.e. we have only doubled directions in which there are isometries, then the right-hand side of the above is zero. In this case if the matrix $\eta_{MN} + P_{MN}$ is invertible, as will be the case for the backgrounds considered in this paper, then the winding must in fact be constant, $\dot{w}^M = 0$. This then implies that integrated equations of motion give exactly
\be
\eta_{MN} \dot{X}^N = \mathcal{H}_{MN} X^{\prime N} \,,
\label{eq:dualityeom}
\ee
which is the correct duality relation for the doubled string \cite{Tseytlin:1990nb, Tseytlin:1990va} and shows that our doubled model indeed reproduces the correct physics of T-duality. When the generalised metric depends on a coordinate the result will generically be non-local duality relations in place of \eqref{eq:dualityeom} as well as non-constant generalised winding. 

\subsection{Mode expansions and equations of motion}

We will now use the action \eqref{eq:doubledaction} to study the three-torus with H-flux. For now consider the four-dimensional doubled space with coordinates $X,Y, \tilde X$ and $\tilde Y$. The monodromy matrix and generalised winding are
\be
P^M{}_N = \begin{pmatrix} 1 & 0 & 0 & 0 \\ 
0 & 1 & 0 & 0 \\
 0 & 2 \pi H N^3 & 1 & 0 \\
 -2\pi H N^3 & 0 & 0 & 1
\end{pmatrix} \quad , \quad w^M = \begin{pmatrix} N^1 \\ N^2 \\ p_1 \\ p_2 \end{pmatrix} \,.
\ee
We therefore need mode expansions for the coordinates and their duals obeying
\be
X(\sigma+2\pi) = X(\sigma) + 2\pi N^1 \quad,\quad
Y(\sigma+2\pi) = Y(\sigma) + 2\pi N^2 \,,
\ee
\be
\tX (\sigma + 2\pi) = \tX(\sigma) + 2 \pi p_1 + 2\pi H N^3 Y(\sigma) \quad,\quad
\tilde Y (\sigma + 2\pi) = \tilde Y(\sigma) + 2 \pi p_2 - 2 \pi H N^3  X(\sigma)\,. 
\ee
These expansions are provided by
\be
X  = x + N^1 \sigma + \sumn \alpha_n e^{in\sigma} \quad,\quad
Y = y + N^2 \sigma + \sumn \beta_n e^{in \sigma} \,,
\label{eq:doubledmodeexpansions0}
\ee
and
\be
\tX = \tilde x + p_1 \sigma + \sumn e^{in\sigma} \talpha_n + HN^3 ( y + \sumn \beta_n e^{in\sigma} )
\sigma + \frac{1}{2} H N^3 N^2 \sigma ( \sigma - 2\pi)\,,
\label{eq:doubledmodeexpansions1}
\ee
\be
\tY = \tilde y + p_2 \sigma + \sumn e^{in\sigma} \tbeta_n - HN^3 ( x + \sumn \alpha_n e^{in\sigma} ) 
\sigma - \frac{1}{2} H N^3 N^1 \sigma ( \sigma - 2 \pi) \,.
\label{eq:doubledmodeexpansions2}
\ee
The quantities $p_1, p_2$ and $\talpha_n,\tbeta_n$ will be determined from the equations of motion. To derive these we insert the above mode expansions into the action \eqref{eq:doubledaction} and carry out the $\sigma$ integration. Note that the generalised metric for this background determines the Hamiltonian term to be explicitly
\be
- \frac{1}{2} \mathcal{H}_{MN} X^{\prime M} X^{\prime N} =  - \frac{1}{2} ( 1 + H^2 Z^2 ) ( X^{\prime 2} + Y^{\prime 2} ) 
- H Z ( X^\prime \tY^\prime - \tX^\prime Y^\prime ) 
- \frac{1}{2} ( \tX^{\prime 2} + \tY^{\prime 2} ) \,.
\ee
The result of the $\sigma$ integration is that the action can be expressed as
\be
S_{doubled} = S_{symplectic} + S_{zero,winding} + S_{\talpha,\tbeta} + S_{\alpha,\beta} \,,
\ee
where the symplectic terms are
\be
\begin{split}
S_{symplectic} = \int d\tau \Bigg[
 &  \dot{x} ( 2 \pi p_1 + \pi HN^3 y - 2\pi^2 H N^3 N^2 )
 +  \dot{y} ( 2 \pi p_2 - \pi HN^3 x + 2\pi^2 H N^3 N^1 ) 
\\ & + 2 \pi \dot{\tilde x} N^1 + 2 \pi\dot{\tilde y} N^2 
+ \frac{2\pi^3}{3} HN^3 ( \dot{N}^1 N^2 - \dot{N}^2 N^1 ) \\
\\
& + 2 \pi  \sumn \dot{\alpha}_n \left( -in \talpha_{-n} + \frac{1}{2} HN^3\beta_{-n} \right)
 + 2 \pi \sumn \dot{\beta}_n \left( -in \tbeta_{-n} - \frac{1}{2} HN^3 \alpha_{-n} \right)\Bigg]\,,
\label{eq:modifiedsympl}
\end{split}
\ee 
the part of the action involving the zero modes and the winding is
\be\begin{split}
S_{zero, winding} & = 
\int d\tau \Bigg[
- \pi (p_1 + HN^3y)^2  - \pi ( p_2 - HN^3x)^2  \\
& - 2\pi (p_2 - H N^3 x) \left( H z N^1 + \pi H N^3 N^1  + H\sumn in \gamma_{-n} \alpha_n  \right)  \\
& + 2\pi (p_1 + H N^3 y) \left( H z N^2 + \pi H N^3 N^2  + H\sumn in \gamma_{-n} \beta_n  \right) 
\Bigg] \,,
\end{split} 
\ee
and that involving the ``dual oscillators'' $\talpha$ and $\tbeta$ is
\be
\begin{split}
S_{\talpha, \tbeta}  = 
2 \pi \int d\tau & \Bigg( 
\sumn \talpha_n \left( - \frac{1}{2} n^2 \talpha_{-n} 
+ H ( z n^2 - N^3 in)  \beta_{-n} + HN^2 in \gamma_{-n} -H \sum_{m \neq 0} mn \gamma_{-m-n} \beta_m 
 \right)
\\
& + 
\sumn \tbeta_n \left( - \frac{1}{2} n^2 \tbeta_{-n} 
- H ( z n^2 - N^3 in)  \alpha_{-n} - HN^1 in \gamma_{-n} + H \sum_{m \neq 0} mn \gamma_{-m-n} \alpha_m 
\right) \Bigg) \,.
\end{split}
\ee
We do not need the explicit form of $S_{\alpha, \beta}$ in what follows. 

We can now work out the equations of motion following from the doubled string with dynamical winding. First of all, 
the equations of motions for the zero modes $\tilde x$ and $\tilde y$ imply that 
\be
\dot{N}^1 = 0 = \dot{N}^2 \,,
\ee
so we must have only constant winding, as expected. If we vary $N^1$ and $N^2$ then we obtain
equations determining $\dot{\tilde x}$ and $\dot{\tilde y}$, which has no effect on the physics. Varying with respect to the momenta (or dual winding) $p_1$ and $p_2$ we find that 
\be
\delta p_1 \Rightarrow  \dot{x}   
-  H N^3 y +  H z N^2 +  \pi H N^3 N^2 +  H \sumn in \gamma_{-n} \beta_n  
=   p_1\,,
\ee
\be
\delta p_2 \Rightarrow \dot{y}
+  H N^3 x -  H z N^1 -  \pi H N^3 N^1 -  H \sumn in \gamma_{-n} \alpha_n  
=   p_2\,.
\ee
Using these equations in the equations of motion resulting from varying the zero modes $x$ and $y$ we learn that in fact also 
\be
\dot{p}_1 = 0 = \dot{p}_2 \,.
\ee
Finally, for the dual oscillator modes $\talpha$ and $\tbeta$ we have
\be
\talpha_n = \frac{1}{in} \dot{\alpha}_n + H  \left(z - \frac{N^3}{in}\right) \beta_n + H \frac{1}{in} N^2 \gamma_n 
 + H \sum \frac{m}{n} \beta_m \gamma_{n-m} \,,
\ee
\be
\tbeta_n = \frac{1}{in} \dot{\beta}_n - H \left(z - \frac{N^3}{in} \right) \alpha_n - H \frac{1}{in} N^1 \gamma_n 
N^3 \alpha_n 
- H \sum \frac{m}{n} \alpha_m \gamma_{n-m} \,.
\ee
If we insert these expressions as well as those for $p_1, p_2$ into the mode expansions \eqref{eq:doubledmodeexpansions1} and
\eqref{eq:doubledmodeexpansions2} we find that we exactly recover the expressions for $\tilde X$ and
$\tilde Y$ that are found by integrating the momentum $P_X$ and $P_Y$, \eqref{eq:tXConstr} and \eqref{eq:tYConstr}.

After solving the equations of motion as above we can 
identify the quantities $\tilde X^\prime$ and $\tilde Y^\prime$ appearing in the doubled
action as the momentum $P_X$ and $P_Y$. We thus have the same Hamiltonian as in the ordinary string
action. In addition we know that all the winding are constant. Then the action of the doubled model is
\be
\begin{split}
S_{doubled}  & = \int d\tau d\sigma \left( \frac{1}{2} \dot{X} P_X + \frac{1}{2} \dot{Y} P_Y 
+ \frac{1}{2} X^\prime \dot{\tX} + \frac{1}{2} Y^\prime \dot{\tY} - \frac{1}{2} \mathcal{H}_{MN} \mathcal{Z}^M \mathcal{Z}^N \right)
\\  &\quad + \pi \int d \tau w^M \eta_{MN} P^N{}_P  \dot{X}^N(0)  
\\ & = \int d\tau d\sigma \left( \dot{X} P_X + \dot{Y} P_Y  - \frac{1}{2} \mathcal{H}_{MN} \mathcal{Z}^M \mathcal{Z}^N + \frac{d}{d\sigma} \left( X \dot{\tX} + Y \dot{\tY} \right) \right) 
\\ & = S_{single} + 2\pi\int d\tau H N^3  X(0) \dot{Y}(0) \,.
\label{eq:doubletosingle}
\end{split} 
\ee
having integrated by parts and discarded total $\tau$ derivatives. The extra term appearing here is
essentially
Hull's ``topological'' term \cite{Hull:2006va}
\be
\frac{1}{2} \int \left( \dot{X} \tX^\prime - X^\prime \dot{\tX} + \dot{Y} \tY^\prime - Y^\prime \dot{\tY} \right) \,.
\ee
We can also understand its appearance as follows. In deriving the equations of motion from $S_{doubled}$ we assumed from the start that the coordinates were not periodic. This was necessary as we wanted to have dynamical winding on the same footing as the ordinary momenta. In this case one finds that $\delta S_{single}$ gives rise to the standard string equations of motion \eqref{eq:singlestringeom} plus an additional term $ - 2\pi \int d\tau H N^3 ( \delta X(0) \dot{Y}(0) - \delta Y(0) \dot{X}(0))$, resulting from the sigma integration by parts if $Z$ is assumed here to wind. This exactly cancels the contribution from the extra term in \eqref{eq:doubletosingle}. Thus we have that $\delta S_{doubled} = \delta S_{single}|_{\mathrm{no}\,\,\mathrm{winding}}$, so that the equations of motion of the doubled model are equivalent to the standard equations of motion of the usual Polyakov action, \eqref{eq:singlestringeom}, as expected from the general discussion in section \ref{sec:actionderivation}.

\section{Non-geometry and non-commutativity}

\subsection{Derivation of Dirac brackets}

We now study the Poisson brackets arising from the symplectic terms of equation
\eqref{eq:modifiedsympl}. In the doubled picture the ``momenta'' such as $p_1$ and $p_2$ are to be
treated as coordinates. This leads to second-class constraints when we come to define the momenta of all
variables in the model. For each variable $\chi$ we define the conjugate momentum $\Pi_\chi$, and have constraints
\be
C_\chi \equiv \Pi_\chi - \frac{\partial L}{\partial \dot{\chi}} \,,
\ee
with initially the standard Poisson brackets
\be
\{ \chi, \Pi_\chi \} = 1 \,.
\ee
Starting with the zero modes only, the second-class constraints are
\be
\begin{split}
 C_x & = \Pi_x - 2 \pi p_1 - \pi HN^3 y + 2 \pi^2 H N^3 N^2 \,,\\
 C_y & = \Pi_y - 2 \pi p_2 + \pi HN^3 x - 2 \pi^2 H N^3 N^1 \,,\\
 C_{p_1} & = \Pi_{p_1} \,,\\
 C_{p_2} & = \Pi_{p_2}\,, \\
 C_{\tilde x} & = \Pi_{\tilde x} - 2 \pi N^1\,, \\
 C_{\tilde y} & = \Pi_{\tilde y} - 2 \pi N^2 \,,\\
 C_{N^1} & = \Pi_{N^1} - \frac{2\pi^3}{3} HN^3N^2\,, \\
 C_{N^2} & = \Pi_{N^2} + \frac{2\pi^3}{3} HN^3 N^1\,.
\end{split}
\label{eq:Czerowind}
\ee
The matrix of Poisson brackets of second-class constraints for the zero modes is $M_{\chi \chi^\prime} = \lbrace C_\chi , C_{\chi^\prime} \rbrace$ for $\chi = \{ x,y,p_1,p_2,\tilde x,\tilde y, N^1,
N^2\}$, and works out as 
\be
M_{\chi \chi^\prime} = 
2 \pi
\begin{pmatrix} 
 0 & - H N^3 & - 1 & 0 & 0 & 0 & 0 & \pi H N^3 \\
 + HN^3 &0  & 0 & -1 & 0 & 0 & - \pi H N^3 & 0 \\
 1 & 0 &0 & 0 & 0 & 0 & 0 & 0 \\
 0 & 1 &0 & 0 & 0 & 0 & 0 & 0 \\
 0 & 0 &0 & 0 & 0 & 0 & -1 & 0 \\
 0 & 0 &0 & 0 & 0 & 0 & 0 & -1 \\
 0 & \pi H N^3 &0 & 0 & 1 & 0 & 0 & - \frac{2\pi^2}{3} H N^3 \\
 -\pi H N^3 & 0 &0 & 0 & 0 & 1 &  \frac{2\pi^2}{3} H N^3 & 0
\end{pmatrix}\,.
\ee
The inverse is
\be
M^{-1}_{\chi \chi^\prime} = \frac{1}{2\pi}
\begin{pmatrix} 
 0 & 0 & 1 & 0 & 0 & 0 & 0 & 0 \\
 0 & 0 & 0 & 1 & 0 & 0 & 0 & 0 \\
 -1 & 0 & 0 & -HN^3  & 0 & -HN^3\pi & 0 & 0 \\
 0 & -1 & H N^3 & 0 & HN^3\pi & 0 & 0 & 0 \\
 0 & 0 & 0 & -HN^3 \pi & 0 & -\frac{2\pi^2}{3} HN^3 & 1 & 0 \\
 0 & 0 & HN^3 \pi & 0 & \frac{2\pi^2}{3} HN^3 & 0 & 0 & 1  \\
 0 & 0 & 0 & 0 & - 1 & 0 & 0 & 0 \\
 0 & 0 & 0 & 0 & 0 & -1 & 0 & 0 \\
\end{pmatrix} \,.
\ee
From this one can read off the non-zero Dirac brackets, which are defined by
\be
\{ f,g \}^* = \{ f, g \} - \{f , C_\chi\} M^{-1}_{\chi \chi^\prime} \{ C_{\chi^\prime}, g\} \,,
\ee
so that we obtain 
\be
\{ x , p_1 \}^* = 
\{ y , p_2 \}^* = 
\{ \tilde x , N^1 \}^* = 
\{ \tilde y , N^2 \}^* = \frac{1}{2\pi}\,,
\ee
\be
\{ p_1 , p_2 \}^* = - \frac{HN^3}{2\pi} 
\quad ,\quad
\{ p_1, \tilde y\}^* = - \frac{HN^3}{2} = \{ \tilde x, p_2
\}^*
\quad,\quad \{ \tilde x , \tilde y \}^* = - \frac{\pi}{3} H N^3\,.
\ee
Meanwhile, for the oscillators we have
\be
\begin{split}
C_{\alpha_n} & = \Pi_{\alpha_n} + 2\pi i n \tilde \alpha_{-n} - \pi H N^3 \beta_{-n} \,,\\
C_{\beta_n} & = \Pi_{\beta_n} + 2\pi i n \tilde \beta_{-n} + \pi H N^3 \alpha_{-n} \,,\\
C_{\tilde \alpha_n} & = \Pi_{\tilde \alpha_n}\,, \\
C_{\tilde \beta_n} & = \Pi_{\tilde \beta_n} \,,
\label{eq:Cosc}
\end{split}
\ee
for which the matrix of Poisson brackets is (with now $\chi = \{ \alpha_n, \beta_n ,\talpha_n, \tbeta_n \})$
\be
(M{}^{osc})_{\chi \chi^\prime} = 2\pi
\begin{pmatrix}
0 & - HN^3 \delta_{n,-m} & in \delta_{n,-m} & 0 \\
 HN^3 \delta_{n,-m} & 0 & 0 &  in \delta_{n,-m}  \\
-in \delta_{n,-m} & 0 & 0 & 0 \\
0 & -in \delta_{n,-m} & 0 & 0 
\end{pmatrix}\,,
\ee
with
\be
(M^{osc})^{-1}_{\chi \chi^\prime} = \frac{1}{2\pi} 
\begin{pmatrix}
0 & 0 & - \frac{1}{in} \delta_{n,-m} & 0  \\
0 & 0 & 0 &  - \frac{1}{in} \delta_{n,-m}  \\
 \frac{1}{in} \delta_{n,-m} & 0 & 0 & - \frac{HN^3}{n^2} \delta_{n,-m} \\
 0 & \frac{1}{in} \delta_{n,-m} &  \frac{HN^3}{n^2} \delta_{n,-m} & 0 \\
\end{pmatrix}\,,
\ee
so that
\be
\{ \alpha_n , \tilde\alpha_m \}^* = 
\{ \beta_n , \tilde\beta_m \}^* = - \frac{1}{2\pi i n} \delta_{n+m} 
\quad,\quad
\{ \tilde \alpha_n, \tilde \beta_m \}^* = - \frac{HN^3}{2\pi n^2} \delta_{n+m}\,.
\ee

\subsection{Non-commutativity}

We can now use these Dirac brackets with the mode expansions
 \eqref{eq:doubledmodeexpansions0}, \eqref{eq:doubledmodeexpansions1},
\eqref{eq:doubledmodeexpansions2} to determine the brackets between our coordinates. 
For each coordinate and its dual we find as expected 
\be
\{ X (\sigma), \tilde X(\sigma^\prime) \}^* = \{ Y(\sigma), \tilde Y(\sigma^\prime) \}^* = - \frac{1}{2\pi} \left( \sigma -\sigma^\prime + \sumn \frac{1}{in} e^{in
(\sigma- \sigma^\prime)} \right) \equiv - \epsilon(\sigma-\sigma^\prime) \,,
\ee
which is compatible with the relationship $\tilde X^\prime = P_X$ between dual coordinates and
momenta. 

In addition we find a non-zero bracket between $\tilde X$ and $\tilde Y$:
\be
\{\tilde X(\sigma), \tilde Y(\sigma^\prime)\}^*= - \frac{\pi}{3} HN^3 - \frac{1}{4\pi} HN^3 ( \sigma - \sigma^\prime)^2 -
\frac{HN^3}{2\pi i} (\sigma - \sigma^\prime) \sumn e^{in(\sigma-\sigma^\prime)} \frac{1}{n}
- \frac{HN^3}{2\pi} \sumn e^{in (\sigma-\sigma^\prime)} \frac{1}{n^2} \,.
\label{eq:noncommbracket}
\ee
These are the coordinates on the non-geometric background and we interpret this result as
non-commutativity in this background. Observe that it is proportional both to the H-flux $H$ and to
the winding $N^3$ about the $Z$-direction. The latter reflects the fact that this non-commutativity
is a consequence of the global properties of the background, with a string wound in the
$Z$-direction feeling the effects of the global patching. 

The bracket \eqref{eq:noncommbracket} is in exact agreement with the result of \cite{Andriot:2012vb} up to different choices of $\alpha^\prime$, and up to the overall constant term 
$-\pi HN^3/3$ resulting from the $\{
\tilde x, \tilde y \}^*$ bracket. This discrepancy is not a complete surprise as in \cite{Andriot:2012vb} the bracket between these zero modes was not fixed by the T-duality rules but instead argued for indirectly. In our case we may note that the origin of this bracket can be traced uniquely to the presence of the term
\be
+ \frac{2\pi^3}{3} HN^3 ( \dot{N}^1 N^2 - \dot{N}^2 N^1 ) =
-\frac{\pi^2}{3} P^M{}_N \dot{w}^N\eta_{MP} w^P + \mathrm{total}\,\, \mathrm{derivative}\,.
\label{eq:thisterm}
\ee
in the Lagrangian. We observe that adding any multiple of $\int d\tau P^M{}_N \dot{w}^N \eta_{MP}
w^P$ to the action does not alter any of our previous results about agreement with the standard
sigma model. Doing so would only affect the $N^1$ and $N^2$ equations of motion, with the result of
shifting the expressions for $\dot{\tilde x}$ and $\dot{\tilde y}$ by terms proportional to
$\dot{N}^1$ and $\dot{N}^2$, which are both zero. Hence in principle we could modify our original
action to remove or modify the term \eqref{eq:thisterm}. It would be interesting to have a precise
check on the presence of this constant term: it is expected that the bracket of
non-commutativity is related to the non-geometric fluxes
\cite{Blumenhagen:2010hj,Lust:2010iy, Blumenhagen:2011ph,Condeescu:2012sp, Mylonas:2012pg, Andriot:2012vb, Bakas:2013jwa}
and perhaps there is a direct argument
from a more (generalised) geometrical point of view (perhaps a derivation of the sigma model from an
underlying geometrical principle such as in \cite{Lee:2013hma}, or some relationship to a geometric
or flux formulation of double field theory \cite{Geissbuhler:2013uka, Berman:2013uda}). 


\subsection{Comment on section condition}

In the Hamiltonian formalism of the bosonic string one has a pair of first class constraints which
generate worldsheet parameterisations. Under Poisson brackets these constraints form a closed
algebra. In the doubled formalism these constraints are
\be
H_1 = \frac{1}{4} (\mathcal{H}_{MN} - \eta_{MN}  ) X^{\prime M} X^{\prime N} \quad , \quad
H_2 = \frac{1}{4} (\mathcal{H}_{MN} + \eta_{MN}  ) X^{\prime M} X^{\prime N}  \,.
\ee
Given a generalised metric $\mathcal{H}_{MN}$ which is an arbitrary function of the doubled
coordinates $X^M$ then the algebra of these constraints does not necessarily close, leading to a
loss of worldsheet diffeomorphism invariance. Thus, as shown in \cite{Blair:2013noa}, one is led to
impose a condition on the background generalised metric. 

In \cite{Blair:2013noa} only backgrounds with trivial $O(D,D)$ monodromy were considered. The Dirac
brackets were $\{X^M(\sigma), X^N(\sigma) \}^* = -\eta^{MN}  \epsilon(\sigma-\sigma^\prime)$, and
worldsheet diffeomorphism invariance as manifested by the closure of the algebra of constraints
required the vanishing of a term involving the Dirac bracket of the generalised metric with itself:
\be
\{ \mathcal{H}_{PQ} ( \sigma) , \mathcal{H}_{RS} (\sigma^\prime) \}^* = \eta^{MN} \partial_M
\mathcal{H}_{PQ} (\sigma) \partial_N \mathcal{H}_{RS} ( \sigma^\prime)  \,.
\ee
This can be implemented by requiring the section condition of double field theory hold on the
generalised metric in the form
\be
\eta^{MN} \partial_M \mathcal{H}_{PQ} \partial_N \mathcal{H}_{RS} = 0 \,.
\ee
Now, for the torus with H-flux we have started with a background that obeys the section condition,
but then gone on to show that we have an additional non-zero Dirac bracket, between two of the dual
coordinates. In our situation this does not modify any of the above concerns, as the generalised metric depends only on the coordinate $Z$ and its brackets receive no modifications due to the H-flux. 

However one might wonder about the general situation. A sufficient condition for closure of the worldsheet diffeomorphism algebra would be
that a background cannot depend on any two coordinates who have a non-zero Dirac bracket, which
ensures that $\{ \mathcal{H}_{PQ}(\sigma), \mathcal{H}_{RS}(\sigma^\prime\}^*=0$. 
The only property of the background that is relevant to determining the bracket is of course the monodromy. 
Ideally one would like to be able to calculate the Dirac brackets for an arbitrary monodromy
$P^M{}_N$, which would allow one to rule out the validity of a given background by checking its
monodromy. Here we were only able to work with a specific example, and check at the end when we had the
brackets that they were consistent with the algebra closure analysis of \cite{Blair:2013noa}. It is
also tempting to speculate about whether one may be able to weaken the section condition for
particular choices of $P^M{}_N$. We leave this for future work.

\section{Non-isometry and no non-associativity} 
\label{sec:noniso}

In this section we allow the non-isometric $Z$ direction to be doubled, allowing us to derive the
Dirac brackets involving the dual coordinate $\tZ$. As the background obtained by T-dualising in all
three directions is not even locally geometric, this leads to new interesting behaviour. The
brackets of $\tZ$ with the other coordinates of this background are non-zero and in fact depend on
the modes of the original string coordinates in an involved fashion. Thus we do not have a geometric
interpretation of the brackets. By taking an additional bracket we can then compute the Jacobi identity,
the failure of which in these backgrounds has been interpreted as evidence for non-associative
behaviour \cite{Blumenhagen:2010hj,Lust:2010iy, Blumenhagen:2011ph,Condeescu:2012sp, Mylonas:2012pg, Andriot:2012vb, Bakas:2013jwa}. 
However, we find that the Jacobi identity is satisfied, and thus our model appears to not see the
non-associative
behaviour. 

\subsection{Dirac brackets for the doubled non-isometry direction}

If the $Z$ direction is doubled then the winding $N^3$ will be treated as dynamical, although we
expect equations of motion to set it to be constant. This puts us in a situation where our monodromy
$P^M{}_N$, which is now meant to be an element of the global $O(3,3)$ group, now involves a
dynamical quantity. However we shall ignore this issue, and think of our boundary conditions as merely involving a set of coordinates and dynamical winding. We can then proceed as before by writing mode expansions for $Z$ and $\tZ$:
\be
Z = z + N^3 \sigma + \sumn \gamma_n e^{in \sigma} \quad,\quad
\tZ = \tilde z+ p_3 \sigma + \sumn \tilde\gamma_n e^{in \sigma} \,.
\ee
Our action is supplemented by the following terms:
\be
\begin{split}
S_Z = \int d\tau & \Bigg[ 
2 \pi \dot{z} p_3 + 2 \pi \dot{\tilde z} N^3 -2\pi \sumn in \dot{\gamma}_n \tilde\gamma_{-n} \\
& + \pi H \dot{N}^3 \Big( 
\sumn 2 \left(  \pi - \frac{1}{in} \right) \left( \alpha_n N^2 - \beta_n N^1 \right) 
+ \sumn ( y \alpha_n - x \beta_n ) \\
& \qquad\qquad + 2 \pi \sumn in \alpha_n \beta_{-n} 
+ \sumn \frac{1}{n} (n-2m) \beta_m \alpha_{n-m} \Big) 
\\
 & - \pi (p_3)^2 - \pi (N^3)^2  - \pi \sumn n^2 ( \tilde\gamma_n \tilde\gamma_{-n} + \gamma_n \gamma_{-n} ) 
\Bigg] \,.
\end{split}
\ee
It is clear that the equations of motion work as expected, ensuring that $N^3$ is constant and $\tZ^\prime = P_Z$. Let us now see what brackets we obtain. Our new second-class constraints are 
\be
\begin{split} 
C_z & = \Pi_z - 2 \pi \dot{z} \,,\\
C_{p_3} & = \Pi_{p_3} \,,\\
C_{\tilde z} & = \Pi_{\tilde z} - 2 \pi N^3\,, \\
C_{N^3} &  = \Pi_{N^3} - \pi H 
\Big(  
\sumn 2 \left(  \pi - \frac{1}{in} \right) \left( \alpha_n N^2 - \beta_n N^1 \right) 
+ \sumn ( y \alpha_n - x \beta_n ) \\
& \qquad\qquad + 2 \pi \sumn in \alpha_n \beta_{-n} 
+ \sumn \frac{1}{n} (n-2m) \beta_m \alpha_{n-m}
  \Big) \,.
\end{split} 
\ee
The non-trivial behaviour here results from the $C_{N^3}$ constraint. This has non-zero Poisson brackets with $C_x, C_y, C_{N^1}, C_{N^2}$, $C_{\alpha_n}$ and $C_{\beta_n}$ of \eqref{eq:Czerowind} and \eqref{eq:Cosc}. The new and important feature is that these Poisson brackets are not constant, but depend directly on the modes. After carefully inverting the matrix of all second-class constraints one finds that this induces the following non-zero Dirac brackets (in addition to the ones we had previously, which are unchanged):
\be
\{ \tilde z , \tilde x\}^*  =  
 \frac{H}{4\pi } \left( 
\frac{2\pi^2}{3} N^2 + \sumn \left( 2 \pi - \frac{2}{in} \right) \beta_n
\right)
 \quad ,
\quad 
\{ \tilde z , \tilde y\}^*  =  
- \frac{H}{4\pi } \left( 
\frac{2\pi^2}{3} N^1 + \sumn \left( 2 \pi - \frac{2}{in} \right) \alpha_n
\right)
\,,
\ee
\be
\{ \tilde z , p_1 \}^*  =  -\frac{H}{4\pi} \left(  y - 2 \pi  N^2 + \sumn\beta_n\right) 
\quad , \quad
\{ \tilde z , p_2\}^*  =  \frac{H}{4\pi} \left(  x - 2 \pi  N^1 +  \sumn\alpha_n\right) \,,
\ee
\be
\{ \tilde z , \talpha_n\}^*  =
 \frac{H}{4\pi in} \left(y  + \left( 2\pi  + \frac{2}{in}\right) N^2 - \beta_n ( 1 + 2\pi i n)  -\sum_{m\neq
0}\frac{2n+m}{m} \beta_{m+n} \right) \,,
\ee
\be
\{ \tilde z , \tbeta_n\}^*  =
- \frac{H}{4\pi in} \left(x  + \left( 2\pi +  \frac{2}{in}\right) N^1- \alpha_n ( 1 + 2\pi i n)  -\sum_{m\neq
0}\frac{2n+m}{m} \alpha_{m+n} \right) \,,
\ee
\be
\{ \tilde z, N^3 \}^*  = \frac{1}{2\pi}
\quad , \quad
\{  z, p_3 \}^*  = \frac{1}{2\pi}\,. 
\ee
That the Dirac brackets of the coordinate $\tZ$ with $\tX$ and $\tY$ involve only the zero mode
$\tilde z$ is consistent with $\tZ^\prime$ being the momentum $P_Z$ conjugate to $Z$, which must
have the usual Dirac brackets. 

The above Dirac brackets guarantee the existence of non-zero Dirac brackets between the coordinate $\tilde Z$ and both of $\tilde X$ and $\tilde Y$. These brackets depend in an involved way on the different modes and it is not clear if they have any geometric interpretation. This fits in with the general expectation that the space we are now considering cannot even locally be described geometrically.

\subsection{Non-associativity?}

Following \cite{Lust:2010iy,Blumenhagen:2010hj,Blumenhagen:2011ph,Condeescu:2012sp, Mylonas:2012pg, Andriot:2012vb, Bakas:2013jwa} 
we can however study the possible non-associativity of these brackets. Although a single bracket of $\tilde Z$ with $\tilde X$ or $\tilde Y$ has an involved dependence on the modes, the bracket of such a bracket has a simpler structure. In particular, one has
\be
\begin{split}
\{ \{ \tilde{Z}(\sigma^{\prime \prime} ) , \tX(\sigma) \}^* , \tY(\sigma^\prime) \}^* & = 
- \frac{H}{12} - \frac{H}{8\pi^2} ( \sigma^2 - \sigma \sigma^\prime) \\
& + H \sumn \frac{1}{4\pi^2 n^2} \left( e^{in\sigma}-e^{in\sigma^\prime} \right) 
+ H \sumn \frac{1}{8\pi^2 in } \left( \sigma^\prime e^{in\sigma} - \sigma e^{in\sigma^\prime} \right)\\
& - H \sumn \frac{1}{4\pi in} \left( e^{in\sigma} - e^{in\sigma^\prime} \right) \\ 
& - H \sumn \frac{1}{4\pi^2 in}  e^{in(\sigma-\sigma^\prime)} \sigma  -H \sumn \frac{1}{8\pi^2n^2} e^{in(\sigma-\sigma^\prime)} + H \sumn \frac{1}{4\pi in} e^{in(\sigma-\sigma^\prime)}\\
& + H \sumn \sum_{\substack{ m\neq0 \\  m\neq -n}} e^{in\sigma}e^{im\sigma^\prime} \frac{1}{8\pi^2}
\frac{1}{nm} \frac{n-m}{n+m} \,.
\end{split}
\ee
Using also the result \eqref{eq:noncommbracket} to get
\be
\{ \tX(\sigma), \tY(\sigma^\prime) \}^*, \tilde{Z}(\sigma^{\prime \prime}) \}^* 
= \frac{H}{6} + \frac{H}{8\pi^2} (\sigma-\sigma^\prime)^2 + H(\sigma-\sigma^\prime) \sumn
\frac{1}{4\pi^2in} e^{in(\sigma-\sigma^\prime)} + H \sumn \frac{1}{4\pi^2n^2}
e^{in(\sigma-\sigma^\prime)} 
\ee
we find that the triple bracket in fact vanishes:\footnote{One might also have argued that a
vanishing triple bracket is the expected outcome here, as one would expect that such a bracket
should obey a cyclic symmetry under interchange of the worldsheet coordinate
$\sigma,\sigma^\prime,\sigma^{\prime \prime}$: as our Dirac brackets only ever involve the zero mode
$\tilde z$ we can never generate a $\sigma^{\prime \prime}$ dependence. This leaves zero as the only
sensible result on the right-hand side of \eqref{eq:triplebracket}. This asymmetry is
a result of our choice for the B-field. It is possible to repeat the calculations for a
symmetric gauge choice, $B_{XY} = HZ/3$, $B_{YZ} = HX/3$, $B_{ZX} = HY/3$, in which case the triple bracket is also
zero.} 
\be\begin{split}
\{ \{ \tilde{Z}(\sigma^{\prime \prime} ) , \tX(\sigma) \}^* , \tY(\sigma^\prime)\}^* + \mathrm{cyclic} &  = 
0 \,.
\end{split}
\label{eq:triplebracket}
\ee
Thus we do not find evidence of a non-associative structure.

This might initially appear a surprise. However, there is a very simple underlying reason. Recall the general definition of the Dirac bracket:
\be
\{ f , g \}^* = \{ f , g \} - \{ f , C_A \} M^{AB} \{ C_B, g \} \,,
\ee
where $M^{AB}$ is the inverse of the matrix of constraints, $M_{AB} = \{ C_A, C_B \}$. One can check
that such brackets indeed 
\emph{always obey the Jacobi identity}, even in the cases where $M_{AB}$ depends on the
coordinates/momenta of the theory (assuming that the original Poisson bracket $\{,\}$ also satisfies the Jacobi identity). Thus the result that the Jacobi identity holds for
the doubled coordinates is in fact inevitable in this framework.

This might seem to indicate that there is no evidence for non-associativity, at least in the classical structure based on the usual (phase space) Dirac brackets. However, one can still argue that there may be a subtle way of introducing non-associativity by linking it to the failure of the Bianchi identity for the $B$-field. 

Consider the following simple example of non-associative behaviour (which is discussed for instance
in \cite{Bakas:2013jwa}). Suppose we have a particle in the background of a magnetic monopole
described by a background gauge field $A_i$. Then, although the \emph{canonical}
momenta $\Pi_i$ and coordinates $X^i$ obey the Jacobi identity, if one studies the \emph{physical}
(gauge invariant) momenta $\mathcal{P}_i = \Pi_i - A_i$ then one easily sees that the triple bracket
$[[\mathcal{P}_i,\mathcal{P}_j],\mathcal{P}_k]+$cyclic of three physical momenta is
in fact non-zero, owing to the failure of the Bianchi identity in the presence of the magnetic
monopole.

The string theory analogue would involve a string in the presence of an NS5 brane. 
This is the background described in the appendix, for which the
three-torus with H-flux is a toy model. 

We may thus speculate that the emergence of non-associative behaviour for the string may be
connected to the failure of the Bianchi identity for the three-form field strength. One may
immediately note that in the three-torus model we have considered it seems impossible to ever see
such a failure, as the Bianchi identity involves an antisymmetrisation on four (spacetime, not dual)
indices, and this model is of course three-dimensional. However, in more realistic configurations it
will be possible. 

The essential point here is that it is necessary to take into account the possible failure of
the Bianchi identity. By working with an explicit local gauge choice for the $B$-field - which is
necessary for the sigma model - it is likely that our method would not see such a failure. 
(Note that the use of the word ``local'' here does not mean the issue arises in the
$Z \rightarrow Z + 2\pi$ patching that is the main concern of the paper. In the NS5 example,
discussed in the appendix, there
is an additional radial coordinate $r$, and one only has a failure of the Bianchi identity at
$r=0$. Thus ``locally'' in this case refers to patches not containing $r=0$.)

This might lead one to suggest defining new coordinates, $\tilde \chi_i$, which are related to the physical momenta of the string as $\tilde \chi_i^\prime =
\mathcal{P}_i$, just as the usual coordinates $\tilde X_i$ are related to the canonical momenta, $\Pi_i$, by $\tilde X_i^\prime =
\Pi_i$. Thus, one would have
\be
\tilde \chi_i^\prime = \tilde X_i^\prime - B_{ij} X^{\prime j} \,.
\label{eq:altcoords}
\ee
which suggests we can take
\be
\tilde \chi_i(\sigma) = \int_{\sigma_0}^\sigma d \tilde \sigma \left( \tilde
X_i^{\prime}(\tilde\sigma) - B_{ij} (\tilde \sigma) X^{\prime j} (\tilde \sigma) \right) \,.
\ee
(There is a zero mode ambiguity in defining $\tilde \chi_i(\sigma)$ which we have chosen to partially fix in
order to have a gauge invariant coordinate.) One can then calculate
\be
\{ 
\tilde \chi_i ( \sigma ) ,
\tilde \chi_j ( \sigma^\prime ) 
\}^*
= \{ \tilde X_i ( \sigma) , \tilde X_j (\sigma^\prime) \}^* +
\int_{\sigma_0}^{\sigma} d \sigma_1
\int_{\sigma_0}^{\sigma^\prime} d \sigma_2
\delta(\sigma_1 - \sigma_2) X^{\prime l} H_{ijl} (\sigma_1) \,,
\ee
after judicious manipulation of various terms involving derivatives of the delta function. Similar
manipulations lead to 
\be
\{
\{ 
\tilde \chi_i ( \sigma ) ,
\tilde \chi_j ( \sigma^\prime ) 
\}^*,
\tilde \chi_k ( \sigma^{\prime\prime} ) 
\}^* 
+ \mathrm{cyclic} 
= 
4 \int_{\sigma_0}^{\sigma} d \sigma_1
\int_{\sigma_0}^{\sigma^\prime} d \sigma_2
\int_{\sigma_0}^{\sigma^{\prime\prime}} d \sigma_3
\delta(\sigma_1-\sigma_2) \delta(\sigma_2 - \sigma_3 ) 
 X^{ \prime l} \partial_{[i} H_{jkl]} (\sigma_1) \,,
\ee
which vanishes if the Bianchi identity holds. 

In a physically realistic situation, such as the NS5 brane, this would lead to a modified version of
the non-commutative bracket and then indeed a non-associative bracket. The non-associativity would
be only felt at the location where the Bianchi identity breaks down, which is at the position of the
NS5 brane. We show this could work in practice for the NS5 brane in the appendix.

One motivation for considering alternative definitions for the dual coordinates is to attempt to
make contact with the approach of \cite{Blumenhagen:2011ph}, where the three-torus with $H$-flux is studied
from a CFT viewpoint. There, one defines special coordinates which are better suited to the CFT
analysis  
(specifically, that are well-defined CFT operators given by integrals of certain conserved currents). These coordinates may be
seen to give rise to duals roughly of the form \eqref{eq:altcoords}, but with a different numerical
prefactor of the $B$-field term. However, it remains unclear how the non-associativity of these papers - which essentially manifests itself in off-shell scattering amplitudes -
relates to the possible non-associativity that is seen here. It would be interesting to quantise the
doubled sigma model in order to further explore these issues.



More fundamentally, it may be that there are subtle issues 
with defining the doubled string action when one wishes to consider T-dualities along non-isometric
directions. As noted above, the boundary conditions encoded by the monodromy matrix
$P^M{}_N$ explicitly involve the dynamical winding $N^3$, and yet this matrix is supposed to be a
constant T-duality. Resolving these issues may require a deeper understanding of the precise
geometrical nature of our model in the doubled space, and of how finite generalised diffeomorphisms are
seen by the sigma model. 

In summary, the lack of a non-associative bracket involving our dual coordinates is an inevitable
consequence of working with Dirac brackets. However, it is possible that in backgrounds for which
the $B$-field Bianchi identity is violated that one can recover a mild non-associativity when using
redefinitions of the coordinates in which the $B$-field appears explicitly. Such redefinitions may
be relevant for CFT-based approaches, e.g. \cite{Blumenhagen:2010hj,Blumenhagen:2011ph}, and may have some interesting doubled geometric meaning (one can view the definition \eqref{eq:altcoords} as being the result of defining a sort of ``flat'' coordinate using the doubled vielbein in the three-torus background, $\partial \tilde \chi^\alpha = E^\alpha{}_M \partial X^M$, for instance).

\section{Conclusions}

We have written down in this paper a proposed doubled string action \eqref{eq:doubledaction}, which
is a simple generalisation of that in \cite{Blair:2013noa} to describe the string in backgrounds
with non-trivial $O(D,D)$ patching. This action 
reproduces the standard equations of motion and duality relations, and leads straightforwardly to the
Dirac brackets of string coordinates. In particular, for the non-geometric background obtained by
acting on the three-torus with H-flux with two T-dualities we have a non-vanishing Dirac bracket
between the two coordinates
$\tilde X$ and $\tilde Y$,
\be
\{\tilde X(\sigma), \tilde Y(\sigma^\prime)\}^*= - \frac{\pi}{3} HN^3 - \frac{1}{4\pi} HN^3 ( \sigma - \sigma^\prime)^2 -
\frac{HN^3}{2\pi i} (\sigma - \sigma^\prime) \sumn e^{in(\sigma-\sigma^\prime)} \frac{1}{n}
- \frac{HN^3}{2\pi} \sumn e^{in (\sigma-\sigma^\prime)} \frac{1}{n^2} \,.
\label{eq:noncommbracketconcl}
\ee
This is in agreement with the result found in \cite{Andriot:2012vb} by a different method, and
implies that the doubled action \eqref{eq:doubledaction} truly knows about the behaviour of the
string in non-geometric backgrounds. 

In addition we are able to use the doubled formalism to
investigate the brackets involving the coordinate $\tilde Z$ on the background obtained by a further
T-duality along a non-isometry direction. There
however we found that the Jacobi identity was satisfied, so that we find no evidence of
non-associativity. Fundamentally, this is due to the fact that Dirac brackets will always obey the
Jacobi identity. Therefore, any non-associative behaviour must have a different origin. We have
suggested above that non-associativity may be connected to the failure of the Bianchi identity of
the $B$-field, which may manifest itself in certain redefinitions of the (dual) coordinates. We
stress, however, that further study is necessary in order to connect such a possibility to
the appearance of non-associativity in the literature, perhaps by quantising the doubled sigma model and
calculating scattering amplitudes. 


One of the strengths of our approach is that the only information about the background which
determines the Dirac brackets is the monodromy matrix. This means that our results hold for any
background sharing the same monodromy, for instance the interesting exotic $5_2^2$ brane example. Unlike the torus with H-flux this is a true supergravity background. However its global non-geometric properties are determined by the same monodromy as in the case of the simpler model. 

There remain open questions regarding the
interpretation of both the doubled action and the resulting Dirac brackets
in terms of the geometry and topology of the doubled space. 
This may have connections
with other approaches to the doubled sigma model such as \cite{Lee:2013hma}, where a closely related
action can be written down by starting from a simple observation about the consequences of the
section condition in double field theory. It is also clear that non-geometric fluxes play an important
role which does not seem to be fully understood for the sigma model, so it would be worthwhile to
investigate how exactly the doubled string feels their effects. This may involve the flux
formulation of double field theory \cite{Geissbuhler:2013uka} or the similar torsionful geometry of
\cite{Berman:2013uda}.

One would also prefer to
be able to derive the Dirac brackets for an arbitrary monodromy, without having to treat specific
cases individually. This would be important for understanding the relationship to the section
condition, which restricts allowed backgrounds. A related goal would be to study so-called ``truly
non-geometric'' backgrounds, which are not T-dual to anything geometric. 

The quantum theories of the backgrounds studied in this paper could also be treated in this approach.

\section*{Acknowledgements}

I would like to thank especially Emanuel Malek and Alasdair Routh for many helpful discussions 
and also David Berman, Dieter L\"ust, Daniel Mayerson and Malcolm
Perry for other useful comments and discussions. I am grateful to the Yukawa Institute of Theoretical
Physics for hospitality at the ``Exotic Structures of Spacetime'' workshop, where part of this work
was carried out. I am supported by the STFC, the Cambridge Home and European Scholarship Scheme and St John's College, Cambridge. 
\appendix

\section{The $5^2_2$ brane}

\subsection{Duality chain}

The torus with H-flux considered above is a toy model and not a true string theory background. However it is very similar to the case of the $5^2_2$ exotic brane. In this short appendix we review the T-duality chain leading to this brane, in order to show that it shares the same $O(D,D)$ monodromy as the torus with H-flux.

The $5_2^2$ brane was studied extensively in 
\cite{
deBoer:2010ud, deBoer:2012ma
}, and has provided an interesting testing ground for studying ideas of double field theory and related concepts such as non-geometric frames 
\cite{
Hassler:2013wsa, Geissbuhler:2013uka, Andriot:2013xca, Andriot:2014uda 
}. The gauged linear sigma model and worldvolume description of this brane has been discussed in 
\cite{ Kimura:2013fda, Kimura:2013zva, Kimura:2013khz , Chatzistavrakidis:2013jqa, Kimura:2014upa} while the behaviour of a rotating doubled string in this background was studied in \cite{Kikuchi:2012za}.

This brane can be obtained by starting with the NS5 brane solution and
performing two T-dualities. The resulting solution is non-geometric, and in fact has essentially
identical $O(2,2)$ monodromy to the three-torus with H-flux. As in the doubled action
\eqref{eq:doubledaction} the symplectic terms are determined entirely by the monodromy matrix this
means we could carry out an entirely equivalent analysis of this background and conclude that it too
has non-commutative coordinates.\footnote{These
backgrounds are solutions of type II string theory. In \cite{Blair:2013noa} the doubled version of
the RNS string was constructed. However, for the purposes of making comments about
Dirac brackets we need only the kinetic terms for the bosonic coordinates,
which are the same.}

We now give the details of the duality chain leading to this brane, essentially following \cite{deBoer:2012ma} (and also \cite{Geissbuhler:2013uka}, which we note is also especially concerned with the failure of the Bianchi identity). 
The standard NS5 brane solution is \cite{Ortin:2004ms}
\be
\begin{split}
ds^2 & = - dt^2 + d\vec{y}_5{}^2 + f d\vec{x}_4{}^2 \,,\\
 B_6 & = (1- f^{-1}) dt \wedge dy^1 \wedge\dots\wedge dy^5 \,,\\
e^{-2\phi} & = f^{-1} \,,
\end{split}
\label{eq:NS5}
\ee
with the harmonic function $f(|\vec{x}_4|) = 1 + \frac{m}{|\vec{x}_4|^2}$. The solution is charged
magnetically under the usual Kalb-Ramond field, and is given above in terms of the dual field for
which $dB_6 = H_7 \equiv \star e^{-2\phi} H_3$. One has from this that $(H_3)_{ijk} =
\epsilon_{ijk}{}^l \partial_l \ln f$. 

We wish to carry out two T-dualities along the transverse directions. In order to do this we need
isometries, and so we take the $x_1 \equiv X$ and $x_2 \equiv Y$ directions to be compact (of radii
$R_X$ and $R_Y$). We also use polar coordinates for the other two transverse directions, which we
call $r$ and $Z \sim Z+2\pi$. 

We then
smear the brane in the $X$ and $Y$ directions. This can be achieved by arraying centres at intervals of $2\pi
R_X$ and $2\pi R_Y$. The harmonic function then involves an infinite sum, which can be successively
approximated by standard methods \cite{Ortin:2004ms} to become
\be
f \approx h_0 + H \log \frac{\mu}{r} \quad , \quad H \equiv \frac{m}{2\pi R_X R_Y} \,.
\ee
Here $h_0$ is a divergent constant, introduced in the second smearing which results in the
appearance of the logarithm, and $\mu$ is a cutoff scale \cite{deBoer:2012ma}. The approximation will be valid for $r >>
R_X,R_Y$ and will break down for $r \sim \mu$. This is related to the fact that our brane is now of
codimension-2 (having compactified two of the four transverse dimensions). Objects of codimension-2
are not expected to exist as well-behaved standalone objects. Rather, they should exist in
superposition with other codimension-2 objects or in the presence of other branes, which we neglect
for $r <  \mu$. As argued in \cite{deBoer:2012ma} however, exotic branes of codimension-2 should
appear generically in string theory. The idea is then that the single brane configuration here can
be used to learn about the general features of such objects, and may represent the (non-)geometry
near an exotic brane, neglecting other sources. 

The twice-smeared NS5 brane solution can be written as
\be
\begin{split}
ds^2 & = -dt^2 + d\vec{y}_5{}^2 + f \left( dX^2 + dY^2 + r^2 dZ^2 + dr^2 \right) \,,\\
B & = H Z dX \wedge dY \,,\\
e^{-2\phi} & = f^{-1} \,.
\end{split}
\ee
This already looks quite similar (in the $X,Y,Z$ directions) to the simple toy model of the
three-torus with H-flux. In particular, the generalised metric for this configuration (restricting
to $X,Y$) is
\be
\mathcal{H}_{MN} = \begin{pmatrix} 
(f+f^{-1} H^2 Z^2) \mathbb{I}_2 
& \varepsilon f^{-1} H Z 
\\
- \varepsilon f^{-1} H Z  &
f^{-1} \mathbb{I}_2 
\end{pmatrix} \quad , \quad \varepsilon = \begin{pmatrix} 0 & 1 \\ - 1 & 0 \end{pmatrix} \,.
\ee
The monodromy of this matrix as we go around the $Z$ circle is the same as for the three-torus situation:
\be
\mathcal{H}_{MN} (Z+2\pi) = P_M{}^P \mathcal{H}_{PQ} (Z)  P_N{}^Q\,,
\ee
with 
\be
P_M{}^{N} = \begin{pmatrix}
\mathbb{I}_2 & 2\pi H \varepsilon \\ 0 & \mathbb{I}_2
\end{pmatrix}  \,.
\ee
One can now focus on the $X$, $Y$ directions and firstly carry out a T-duality in the $X$ direction to arrive at the Kaluza-Klein monopole, with vanishing $B$-field and dilaton, and metric
\be
ds^2 = - dt^2 + d\vec{y}_5{}^2 +  f^{-1} ( d \tX - H Z dY )^2 + f ( dY^2 + r^2 dZ^2 + dr^2 ) 
\quad , \quad \tX \sim \tX + 2\pi H Y \,.
\ee
The situation here is analogous to that of the twisted torus. A second T-duality, in the $Y$ direction, gives us an exotic background known as the $5^2_2$ brane:
\be
\begin{split}
ds^2 & = -dt^2 + d\vec{y}_5{}^2 + \frac{f}{f^2 + H^2 Z^2} ( d\tX^2 + d\tY^2) + f ( r^2 dZ^2 + dr^2 )
\,,\\ 
B &  = - \frac{H Z}{f^2+H^2 Z^2} d\tX \wedge d\tY \,, \\
e^{-2\phi} & = \frac{f}{f^2 + H^2Z^2} \,.
\end{split} 
\ee
Our work in the rest of this paper then indicates that the non-commutativity result should continue
to hold in the present situation.

Similarly, one could use the doubled picture to perform a T-duality in the direction $Z$, leading
once more to a background which is not locally geometric. 

\subsection{Failure of the Bianchi identity and non-associativity}

For the NS5 brane, the three-form is $H_{ijk} = \epsilon_{ijk}{}^m \partial_m \ln
f$, so that
\be
4 \partial_{[i} H_{jkl]} 
= 
- \bar{\epsilon}_{ijkl} \delta^{mn} \partial_m \partial_n f \,,
\ee
where $\bar{\epsilon}_{ijkl}$ is the Levi-Civita tensor on flat $\mathbb{R}^4$. The harmonic
function is $f=1 + \frac{m}{|\vec{x}_4|^2}$. As a result, one can show that
\be
4 \partial_{[i} H_{jkl]} 
=
- 4 \pi^2 m \delta^{(4)} ( \vec{x} ) 
\,.
\ee
After smearing on the $x_1\equiv X$ and $x_2\equiv Y$ directions, and passing to cylindrical
coordinates, one can show that the failure of the Bianchi identity is
\be
4 \partial_{[X } H_{YZr]} = H \delta(r) \,,
\ee
where $H \equiv \frac{m}{2\pi R_X R_Y}$ is the value of the field strength of the $B$-field in the
smeared form of the solution. 

We can then look at the bracket of the alternative coordinates \eqref{eq:altcoords} (evaluated at
the same worldsheet point)
\be
\begin{split}
\{
\{ \tilde \chi_X(\sigma) , \tilde \chi_Y (\sigma) \}^*,
\tilde \chi_Z (\sigma)
\}^* + \mathrm{cyclic}
& = H 
\int_{\sigma_0}^{\sigma} d \sigma_1
\int_{\sigma_0}^{\sigma} d \sigma_2
\int_{\sigma_0}^{\sigma} d \sigma_3
\delta(\sigma_1-\sigma_2) \delta(\sigma_2 - \sigma_3 ) 
r^\prime(\sigma_1) \delta( r ( \sigma_1 ) ) \\
& = H 
\int_{\sigma_0}^{\sigma^\prime} d \tilde\sigma
r^\prime(\tilde \sigma) \sum_i \frac{\delta(\tilde\sigma - \sigma_i)}{| r^{\prime}( \sigma_i)|} \,,
\end{split} 
\ee
where the sum in the final line is over $\sigma_i$ such that $r(\sigma_i) = 0$. We thus see that we pick up a
contribution of $H$ to a non-associative result depending on whether the string passes through
$r=0$, the position of the brane.

\bibliographystyle{JHEP}
\bibliography{NewBib}

\end{document}